\documentstyle[twocolumn,prl,aps]{revtex}
\widetext

\def\beqn{\begin{eqnarray}}
\def\beqns{\begin{eqnarray*}}
\def\eeqns{\end{eqnarray*}}
\def\beq{\begin{equation}}
\def\eeq{\end{equation}}
\def\bea{\begin{array}}
\def\ea{\end{array}}

\def\<{\langle}
\def\>{\rangle}

\begin{document}

\twocolumn[\hsize\textwidth\columnwidth\hsize\csname @twocolumnfalse\endcsname
\draft
\preprint{}

\title{\sc Investigation of quantum transport by means of O(N) real-space methods}

\author{\sc Stephan ROCHE}

\address{Department of Applied Physics, University of Tokyo, 7-3-1 Hongo,
Bunkyo-ku, Tokyo 113, Japan.}

\maketitle

\begin{abstract}
\leftskip 54.8pt
\rightskip 54.8pt
We investigate quantum transport by a suitable development of the
Kubo formula on a basis of orthogonal polynomials 
and using real space recursion approaches. We show that the method
enable to treat systems close to a metal-insulator transition.
On quantum Hall systems, results are given in the context of 
recent new universal relations between transport coefficients. 
RKKY mesoscopic interaction is also evaluated for
two-dimensional quasiperiodic systems and insight of the usefulness
of our method is given for large scale computational problems.
\end{abstract}


]
\vspace{40pt}

\vspace{20pt}

\section{Introduction}

\vspace{20pt}

\hspace{\parindent}Quantum transport
 in high dimensional non periodic systems, 
is usually investigated by way of scaling analysis through diagonalization of periodic
hamiltonian. However, if N is the number of states, exact
diagonalization requires a CPU time scaling as $O(N^{3})$,
 and memory scaling as $O(N^{2})$. For
sparse Hamiltonians, the use of Lanczos algorithm reduces memory and
CPU time requirements to $O(N)$.~\cite{Hay,ON} In the present work,
 quantum transport in presence of magnetic field using a
development of Kubo formula on orthogonal polynomials, is
investigated. This approach
has been developed in the context of strongly disordered~\cite{Mayou-I},
quasiperiodic~\cite{RO-QP} and inhomogeneous systems where 
electronic susceptibility and conductivity were successfully estimated.
~\cite{RO-RKKY,ROB-RKKY} The key point of the algorithm
 is to rescale the density of states after an evaluation of the upper
and lower bounds on energy and then to make a polynomial expansion of
the associated Kubo formula. In principle, any orthogonal polynomials 
may be used, but it turns out
that manipulations of Chebyshev polynomials are particularly suitable
as they are isomorphic to Fourier series.~\cite{Magnus} Possible
 applications to nanostructures will be mentioned in this context.

\vspace{20pt}

\hspace{\parindent}In two-dimensional systems, 
the study of localization in a magnetic field enables 
to address problems related to the quantum Hall effect and
metal insulator transitions.~\cite{QHE_Klitz} A magnetic field in a pure
system introduces further topological complication of the
 electronic spectrum, which turns out to be a degenerate ensemble of discrete
Landau levels. For disordered systems, in the
limit of strong magnetic field, perturbational, numerical and field
 theoretical approaches have depicted a comprehensive view 
of the corresponding physical phenomena.~\cite{KRAM} In real
materials, due to disorder, 
Landau levels are enlarged and overlap to form Landau bands
but extended states still persist at the center of each 
Landau bands, as revealed for instance by the divergence of the
 localization length near these critical energies.~\cite{Ando}
The studies of the associated eigenstates inferred a complicated nature described by
multifractality.~\cite{Chalker,Huk}

\vspace{20pt}

\hspace{\parindent}The concept of multifractal states,
also proposed by Kohmoto~\cite{Kohmoto-QP} for quasiperiodic systems,
 is an important related issue. Indeed, 
anomalous quantum diffusion in quantum Hall systems~\cite{QHE-AD} (QHS) and  
in quasicrystals~\cite{RO-QP,Herman} may have some direct relations with
the observed physical properties. The possibility of an experimental
observation of multifractal exponents in the QHS has been proposed recently by
Brandes et al.~\cite{Tobias} To understand properly the effects of
these complicated eigenstates, one has to develop suitable methods.

\vspace{10pt}

\hspace{\parindent}In this context, after discussing briefly the spectrum of the
pure systems with magnetic field, we present a new method to
evaluate the quantum transport coefficient in various situations. For
QHS, the method brings complementary results for studying the
transition regions when $\sigma_{xy}$ goes from one quantized plateau
to the next, while $\sigma_{xx}$ goes through a peaked value and
decrease again. It also enables a study of the phase diagram of the integer
quantum Hall effect (QHE) through the calculation
 of the transport coefficient as a function of disorder and magnetic
strengthes. Besides, we show that physical phenomena indirectly related to electronic
propagation, such as RKKY interaction, may be also investigated in
high-dimensional aperiodic systems, by using similar
algorithms. Finally, applications to large-scale computational methods 
is suggested.

\vspace{10pt}

\section{Non interacting electrons
 in 2D systems with magnetic field}

\subsection{Spectral properties for pure systems} 

\vspace{20pt}

\hspace{\parindent}The density of states on 2D disordered systems with
magnetic field is a well established result which can also be
investigated by recursion method.~\cite{MF4} In the zero-disorder limit, 
one gets a Landau level type spectrum where the
number of gaps are defined by the dimensionless measure (magnetic strength)
$\alpha=eBa^{2}/2\pi c$ (a: the lattice unit, B magnetic field). To
investigate spectral and transport properties by recursion method, let's
consider the tight-binding representation of the hamiltonian

$$H=\sum_{n_{x},n_{y}}\varepsilon_{n}|n_{x},n_{y}\rangle
\langle n_{x},n_{y}|+$$

$$
\sum_{n_{x},n_{y}} t (|n_{x}+1,n_{y}\rangle
+ |n_{x}-1,n_{y}\rangle + e^{2i\pi\alpha n_{x}}|n_{x},n_{y}+1\rangle
e^{-2i\pi\alpha n_{x}}|n_{x},n_{y}-1\rangle) \langle|n_{x},n_{y}| $$

\vspace{10pt}

\noindent
A recursive construction of an orthogonal basis 
$|\Psi_{n}\rangle$ such that

\begin{eqnarray}
{\cal H}|\Psi_n\> &=& a_n|\Psi_n\> + b_n|\Psi_{n+1}\> + b_{n-1}|\Psi_{n-1}\>\nonumber\\
a_n &=& \<\Psi_n|{\cal H}|\Psi_n\>\nonumber\\
b_n &=& \<\Psi_{n+1}|{\cal H}|\Psi_{n}\>\nonumber
\end{eqnarray}

\noindent
tridiagonalizes the hamiltonian and enables to evaluate the
spectral properties quickly and accurately \cite{Hay}. In the one-band case, the
series $a_{n},b_{n}$ usually converge after about 20 recursion steps, and the
limit $a_{\infty},b_{\infty}$ give the final form of the Jacobi
matrix, e.g. the convergence of the continuous fraction development of the Green's
function is achieved.

\begin{equation}
{\cal J}=
\left(
\begin{array}{cccccc}
a_{0}& b_{1}  & 0   &\ldots &\ldots &0     \\
b_{1}& a_{1}  &b_{2} & 0   &\ldots &0       \\
0    & b_{2} &a_{2}&\ddots & 0 &\ldots        \\
\vdots& 0 \ddots& \ddots &\ddots& b_{\infty}&\ldots    \\
\vdots & \ldots & 0 & b_{\infty} &a_{\infty} & b_{\infty} \\
0      &     &  \ldots &  0  & b_{\infty} & a_{\infty} \\  
\end{array}
\right)\nonumber
\end{equation}

\noindent
Introducing a magnetic field with rational flux $\alpha=\frac{P}{Q}$
leads to Q bands, separated by gaps. The asymptotic regime is
given by $q$ distincts values $b_{q}, q=1\ldots,Q$, moreover the terminator 
of the continuous fraction expansion can be evaluated exactly :

$$
{\cal T}(z) = \frac{1}{\displaystyle z - \frac {\strut b_0^2}
{z  - \frac{\displaystyle \strut b_1^2} 
{\frac {\displaystyle \ddots} {\displaystyle \strut z  -
b_{q}^2{\strut \cal T}(z)}}}}
\ \ \ \ \ \
{\cal T}(z)=\frac{\alpha_{q}+\beta_{q}{\cal T}(z)} 
{\gamma_{q}+\delta_{q}{\cal T}(z)}
$$

\noindent
with

\begin{eqnarray}
\gamma_{n}&=&z\gamma_{n-1}-b_{q-n+1}\alpha_{n-1},
\alpha_{n}=\gamma_{n-1}\nonumber\\
\delta_{n}&=&z\gamma_{n+1}-b_{q-n+1}\beta_{n-1},
\beta_{n}=\delta_{n-1}\nonumber
\end{eqnarray}

\noindent
For $n=2,\ldots, q$ and initial values $\alpha_{1}=1,\beta_{1}=0,
\gamma_{1}=z, \delta_{1}=-b_{q}^2$. 
The explicit form of the terminator and  Green's function are thus given by :

$${\cal T}(z)=\frac{1}{2\delta_{q}}
(\beta_{q}-\gamma_{q}\pm
\sqrt{(\beta_{q}-\gamma_{q})^{2}+4\delta_{q}\alpha_{q}} )$$

$$\<\psi_{0}|G(z)|\psi_{0}\> = \frac{1}{\displaystyle z - \frac {\strut b_0^2}
{\displaystyle z - \frac{\strut b_1^2}{z - \frac{\strut b_2^2}
{\strut \displaystyle z - ... \frac{\ }{\ \ \frac{\strut b_{q-1}^2}
{\strut \displaystyle z - b_q^2{\strut \cal T}(z)}}}}}}$$

%%%%%%%%  FIGURE I  %%%%%%%%%%%%%%%%%%%%%%%%%%%%%%%%%%%%%%%

\vspace{10pt}

\noindent
so that the exact density of states can be determined.\cite{Turchi} On
Fig. 1, the recursion coefficients $b_{n}$ and the
corresponding TDoS for $\alpha=0,1/3,1/8$ are presented. The relation between 
the number of bands and the asymptotic behavior of recursion
 coefficients becomes obvious.

\vspace{10pt}

\hspace{\parindent}We now introduce the disorder through 
the site energies, which are chosen randomly (with uniform
probability) within the interval  $[-W/2,W/2]$. The presence of disorder will 
smear out the gaps. We check that for $W=6t$ all the gaps have
disappeared.

\vspace{20pt}

\hspace{\parindent}In the following, after discussing current open
problems of the integer quantum Hall effect, the method we used
 to investigate Kubo formula \cite{Kubo}, different from usual Landauer
method \cite{Landauer,Kinnon}, is presented. It was
initially proposed by Mayou~\cite{Mayou-I} for studying frequency dependent conductivity 
$\sigma(\omega)$ in disordered systems.
 An investigation of Kubo conductivity and anomalous quantum diffusion 
in 3D quasiperiodic systems has been carried out by developing a similar
algorithm for static conductivity at zero temperature.~\cite{RO-QP}

\vspace{10pt}

\subsection{ Effect of disorder on quantum transport in magnetic fields}

\vspace{10pt}

\hspace{\parindent}In the presence of disorder, 
extended states exist only at energies for which localization 
length  diverges as $\xi\sim(E-E_{c})^{-\nu}, \
\nu=2.4\pm0.1$.~\cite{KRAM} At the critical value $E_{c}$,  some controversies about the universality of the absolute value
for the diagonal conductivity still remain. Experimentally $\sigma_{xx}^{c}\sim
0.2-0.5$ which is in agreement with some numerical scaling analysis of
$\sigma_{xx}^{L}(E,\eta)$.~\cite{MF1,MF2,MF3}

\vspace{10pt}

\hspace{\parindent}The question of the disappearance
 of these extended states is important.
 In a study by Yang and Bhatt~\cite{MF3},
a critical strength of randomness $W_{c}$ by which all
the extended states vanish, was found to be independent of
magnetic field ($W_{c}\sim 6.t$, with $t$ the constant hopping term).
 Since, in their calculations, $W_{c}$ still persist in the 
weak magnetic field limit, the driven mechanism of the metal-insulator 
transition was consistent with the floating up
 picture of extended states, as proposed
theoretically by Khmel'nitzkii and
Laughlin.~\cite{Kmel,Laughlin} According to their argument,
 the energies of extended states within a given energy range 
in the vicinity of the center of a landau band 
$E_{n}=(n+1/2)\hbar\omega_{c}\frac{1+(\omega_{c}\tau)^{2}}{ (\omega_{c}\tau)^{2}}$
tend to infinity if magnetic field is vanishingly small (as expected
by Anderson theory of localization).  This 
scenario turns out to be crucial for the global phase diagram proposed by
 Kivelson, Lee and Zhang (KLZ).~\cite{Kivelson} The KLZ approach
leads to a physical relation between different quantum Hall liquids,
known as the law of corresponding states. Accordingly, 
transitions from a quantum Hall states to insulators
are allowed for certain values of $\nu=1,1/3,...$, and forbidden
for others ($\nu=2,3,4...$, $\nu=2/5,...$).

\vspace{10pt}

\hspace{\parindent}However, this picture has been 
recently contradicted by recent experiments by Song et
al. \cite{Song} where transition from $\nu=2$-state to insulator were observed.
 Sheng and Weng have also reported numerical support
 of a continuous disappearance of IQHE
in a tight-binding model, and related to merging of extended states
separating different plateaus.~\cite{MF4} The authors claim that their 
results in the weak magnetic field limit are not consistent with the
 floating up picture. In addition, they found that at the metal-insulator transition
dissipative and non-dissipative conductivities are equal 
$\sigma_{xx}(E_{c},W_{c})=\sigma_{xy}(E_{c},W_{c})$.~\cite{MF4,Song}
The physical explanation of this phenomenon still remains unclear and
in the following we will show that their numerical calculations can be 
criticized.

%,such that

%\vspace{10pt}

%\begin{eqnarray}
%\sigma_{xx}(\nu)&\leftrightarrow& \sigma_{xx}(\nu+1), \ 
%\sigma_{xy}(\nu)\leftrightarrow
% \sigma_{xy}(\nu+1)-\frac{e^{2}}{h}\nonumber\\
%\sigma_{xx}(\nu)&\leftrightarrow& \sigma_{xx}(1-\nu), \ 
%\sigma_{xy}(\nu)\leftrightarrow \frac{e^{2}}{h}-
%\sigma_{xy}(1-\nu)\nonumber\\
%\rho_{xx}(\frac{\nu}{1+2\nu})&\leftrightarrow
%&\rho_{xx}(\nu) \nonumber\\
%\rho_{xy}(\frac{\nu}{1+2\nu})&\leftrightarrow&
%\rho_{xy}(\nu)+2\frac{h}{e^{2}}
%\nonumber
%\end{eqnarray}

\vspace{10pt}

\subsubsection{Kubo formula by recursion: diagonal conductivity}

\vspace{10pt}

\hspace{\parindent}The real-space calculation of the
diagonal Kubo formula of the electronic
conductivity may be considered as an alternative for usual Landauer
conductance calculations, or diagonalization methods. The use of Landauer formula for
investigating quantum zero temperature transport is usually
associated with free escape boundary conditions. One direction of the 
system is periodic whereas the other, of size $L$, is connected to
metallic leads with different chemical potentials. 
Scaling analysis is performed through $L$. By recursion
method, we avoid exact diagonalization of the Hamiltonian, so
that we can treat in principle larger and more complex systems. To
reduce the possible numerical instability at boundary conditions 
induced by the velocity operator (periodic boundary conditions will
indeed generate a short-circuit across the sample),
 we transform the Kubo formula in the following way
 (${\hat{{\cal X}}(t)}= {e}^{i{\cal
H}t/\hbar}{\hat{{\cal X}}}{e}^{-i{\cal H}t/\hbar}$ and ${\hat{{\cal X}}}$
is the component along direction x of the position operator, $\Omega$
the volume of the system) :

$$
{\displaystyle 
\sigma_{xx}(E_{F})= \frac{2\hbar {e}^{2}\pi}{\Omega} \lim_{t\to\infty}
\hbox{Tr}[\delta(E_{F}-H)\frac{(\hat{\cal X}(t)-\hat{\cal X}(0))^{2}}{t}]}$$

\noindent
and we keep control of the asymptotic behavior of
the quantum diffusion of the wave-packets.~\cite{RO-QP} The
conductivity reads :

$$\frac{2\hbar {e}^{2}\pi}{\Omega} \sum_{j_{x},j_{y}} {\cal D}_{j}(t)\times
{\displaystyle \Im m_{\eta\to 0}} <\widetilde{\Phi}_{j}(t)
 \mid  G (E_{F}+i\eta) \mid \widetilde\Phi_{j}(t)>$$

\noindent
where ${\mid \Phi_{j}(t)>}= \hat{\cal X} e^{-i{\cal H}t/\hbar}{\mid\ j>}$
and $\mid\widetilde\Phi_{j}(t)>$ is normalized. The summation should be
done over the total basis of states $|j_{x},j_{y}\rangle$, but it turns out
that a limited number of initial sites is sufficient to achieve
convergence of the calculation. The time-dependent evolution of a
wave-packet initially localized at $|j_{x},j_{y}\rangle$ is also
evaluated by polynomial expansion of the evolution operator 
$ e^{-i{\cal H}t/\hbar}=\sum_{n}(\int dE {\cal P}_{n}
e^{-i{E}t/\hbar}){\cal P}_{n}({\cal H})$ where we choose Chebyshev
polynomials of first kind (see appendice).\cite{RO-QP}

\vspace{10pt}

\hspace{\parindent}Finally, one could also define a scaling
parameter from the construction of the recursion
basis. Indeed, after computing N recursion steps,
 the energy resolution is roughly $\Gamma\sim
\frac{W}{N}$ where $W$ is the total bandwidth of the DoS. Starting from a
localized state in  $|\Psi_{0}\rangle= |j_{x},j_{y}\rangle$
 the time corresponding to the
``propagation'' of the N states 
$|\Psi_{N}\rangle={\cal P}_{N}|\Psi_{0}\rangle$ is $\tau\sim
\frac{\hbar N}{W}$. Accordingly, a scaling analysis as a function of N
may be done. Besides, the use of Kubo formula implicitly
requires taking the limit of an infinite system, as the spectrum 
of any finite system is discrete. By retaining a finite $\eta$
imaginary part in Green's function, one replaces the
delta function by a peaked smooth function of width $\eta$, which must 
be greater than the level spacing in the finite system. 
The thermodynamic limit is achieved with increasing system size
$L\rightarrow\infty$, and $\eta\rightarrow 0$ in order to retain all
 the contributions from the spectrum. In our calculation, we find that
the thermodynamic limit is achieved for a finite number of initial states
 $|j_{x},j_{y}\rangle$. Concerning the transition regions mentionned
earlier, one notes that the effect of finite temperature or frequency
is to smear the QH-metal-QH phase transitions.

\vspace{10pt}

%%%%%%%%  FIGURE II  %%%%%%%%%%%%%%%%%%%%%%%%%%%%%%%%%%%%%%%

\hspace{\parindent}On Figure 2, the diagonal
conductivities obtained for a disorder and magnetic strength $W=2,5$ 
 $\alpha=1/3$ and $\eta=0.15$, are similar to those from
reference.~\cite{MF3} The increasing of disordered strength leads to
a reduction of the conductivity, as well as a shift
 of extended states towards the center of the band (as
indicated by the arrows on Figure 2).  Ando 
 first described numerically this effect by studying the density of states and 
Thouless numbers.~\cite{Ando-II,Ando-III} A physical interpretation of this levitation
mechanism associated to Landau-level mixing has been proposed for high 
magnetic field.~\cite{Haldane}

\vspace{10pt}

%%%%%%%%  FIGURE II  %%%%%%%%%%%%%%%%%%%%%%%%%%%%%%%%%%%%%%%

\hspace{\parindent}The question of universality of the diagonal
conductance at critical energies is also addressed in Figure 3. As the 
finite imaginary part of Green's function tends to zero, one
clearly sees that the $e^{2}/2h$-limit is approached with our method. One
notes that the figures ($\eta=0.15,0.09,0.05$), in the inset, are obtained for
only one initial state $|j_{x},j_{y}\rangle$, so that the fluctuations 
of the exact shape are artificials. The central figure is an average
result on ten different sites.

%%%%%%%%  FIGURE III  %%%%%%%%%%%%%%%%%%%%%%%%%%%%%%%%%%%%%%%

\vspace{10pt}

\hspace{\parindent}We also discuss the numerical
results obtained in~\cite{MF2} for $\alpha=1/16$ and different values
of the disorder strength. To study the universal relation between
 the transport coefficient, the authors have computed
$\sigma_{xx}(W_{c},E_{F})$ and $\sigma_{xy}(W_{c},E_{F})$ for
$E_{F}=-2.75$ in proper units. On Fig. 4,
$\sigma_{xx}(W,E_{F}=-2.75)$ for different disorder strengthes 
is shown. The Fermi energy $E_{F}=-2.75$  considered in~\cite{MF2}
turns out to lie within a gap of the spectrum. The
 results from~\cite{MF2} are then inconsistent with the physics of the
problem. Indeed, if the Fermi energy is located within a gap of the pure
system, by increasing disorder, a consequent enhancement of the
conductivity may be induced for a finite system, as it is illustrated on
Figure 5. However, if the Fermi level lies within the center of a Landau band, such a
behavior is not seen in our results for the Fermi energy in the center 
of a Landau band. The authors of~\cite{MF2} may have not
consider the correct location of the Fermi energy (remind that the
position of the extended state will be affected by the disorder) and
accordingly, their statement is at least numerically questionnable.

%%%%%%%%  FIGURE IV  %%%%%%%%%%%%%%%%%%%%%%%%%%%%%%%%%%%%%%%

\vspace{10pt}

\hspace{\parindent}On figure 6, we show the $\sigma_{xx}(E_{F})$ for
two different values of disorder. The black curves are the results
obtained for one site $ {\cal D}_{j}(t)\times
\Im m_{\eta\to 0} <\widetilde{\Phi}_{j}(t)
 \mid  G (E_{F}+i\eta) \mid \widetilde\Phi_{j}(t)>$, 
whereas the red one is the averaged result. One 
can see that the fluctuations of the components, entering in
the sum of the Kubo expression in real space, is not critical, so that 
finite number of sites is sufficient to achieve convergence.

%%%%%%%%  FIGURE V  %%%%%%%%%%%%%%%%%%%%%%%%%%%%%%%%%%%%%%%

\vspace{10pt}

\subsubsection{Hall Kubo conductivity}

\vspace{10pt}

\hspace{\parindent}The most spectacular result for 2D electrons in
magnetic field is given by the quantization of the Hall conductance in units of
$\frac{e^{2}}{h}$, when the Fermi level lies within a gap. This result has
been well established and related to topological invariant known as
the Chern numbers, which count the exact number of extended states
up to the Fermi energy. Contribution to electronic conductivity of a given eigenstate
$\mid k>$ can be calculated from 

\vspace{10pt}

\begin{eqnarray}
\sigma_{xy}^{k}&=&\frac{ie^{2}\hbar}{A}\sum_{q \ne k}
\frac{<k|\hat{V}_{y}|q><q|\hat{V}_{x}|k>-<k|\hat{V}_{x}|q><q|\hat{V}_{y}|k>}
{(\varepsilon_{q}-\varepsilon_{k})^{2}} \nonumber \\
<\sigma_{xy}^{k}>&=&\frac{1}{4\pi^{2}}\int d\varphi_{1}d\varphi_{2}
\sigma_{xy}^{k}(\varphi_{1},\varphi_{2})=\frac{e^{2}}{h}{\cal N}_{k}
\nonumber
\end{eqnarray}

\vspace{10pt}

\hspace{\parindent}On finite size systems, Hall
conductance (boundary condition average) 
enables us to identify if $<\sigma_{xy}^{k}>$ is a so-called 
non-zero Chern number ${\cal N}_{k}$ or not,
 thus if the corresponoing $|k\rangle$ is an extended state or
localized state (formulation is due to Thouless, Kohmoto, Nightingale and den
Nijs 's~\cite{TKNN,Kohmoto}). Recently, non-commutative geometry
 has provided an interesting framework to investigate 
Kubo formula and quantum Hall effect.~\cite{Bell-QHE}

\vspace{10pt}

\hspace{\parindent}As we are interested, in particular, in computing the
non-dissipative conductivity for energies in the transition regions,
(where $\sigma_{xy}$ is not quantized), one has to adopt a different
strategy. To that end, we propose a real-space approach of Hall
conductivity. Starting from the general off-diagonal form of the Kubo
conductivity, one shows that the proper algorithm allowing the
expansion of the Hall conductance in a real-space basis is given by :

$$\sigma_{xy}=-\frac{ie^{2}\hbar}{2\Omega}
{\displaystyle \int}_{E_{2}>E_{F}\atop E_{1}<E_{F}}
dE_{1}dE_{2} \frac{f(E_{1})-f(E_{2})}{E_{1}-E_{2}} 
\hbox{Tr} [\delta(E_{1}-{\cal H})\dot{\cal Y}
\delta(E_{2}-{\cal H})\dot {\cal X}]$$

\noindent
enables us to expand the spectral measure on the basis of Chebyshev
polynomials. After some simple algebra, one finds two parts to be
evaluated separately :

$$\sigma_{xy}=-\frac{ie^{2}\hbar}{2\Omega}
\sum_{m,n,i} I_{mn}\times\langle i_{x},i_{y}| {\cal P}_{n}({\cal H})\dot{\cal Y}
 {\cal P}_{m}({\cal H})\dot{\cal X}|i_{x},i_{y}\rangle$$

\noindent
where $I_{mn}$ is analytical and depends on the
choice of the polynomial basis. In our case, it corresponds to
 ($A_{F}=$Arcos$({{E_{F}-a}\over{2b}})$, and $a.b$ associated to the
weight function of Chebyshev polynomials) :

$$
I_{mn}={1\over{\pi b}} \biggl{ \{ }
{{\sin(m+n+3)A_{F}}\over{(m+n+3)}}-{{\sin(m+n+1)A_{F}}\over{(m+n+1)}}
\biggr{ \} }\ $$

% +
% \ {-i\over{b\pi}}{ {(2^{m+n+2}+(-2)^{n+m+2})}\over{(m+n+1)(m+n+3)}}
%$$

\noindent
The other part implies the calculation of the coefficients
 $\langle j_{x},j_{y}|{\cal P}_{n}({\cal H})\dot{\cal Y}
 {\cal P}_{m}({\cal H})\dot{\cal X}| j_{x},j_{y}\rangle$.
A reasonable number of initial sites $|j_{x},j_{y}\rangle$ should be
considered. The sum over m and n-indice is, given the form of the
$I_{mn}$ factors, limited by some appropriate cut-off. One notes that the
computational time is however much larger compared to the
$\sigma_{xx}$ calculation. Further work is in progress for testing
the accuracy as well as the gain over 
conventional diagonalization procedures.~\cite{RO-K}

\vspace{10pt}

\section{RKKY magnetic interaction in non-periodic systems}

\vspace{10pt}

\hspace{\parindent}Amongst interesting phenomena related to electronic 
propagation, the Rudermann Kittel Kasuya Yosida interaction
(RKKY)~\cite{RKKY} between magnetic sites in disordered systems has been subjected 
to great attention. In particular it has been shown to be very
important for understanding the spin glass transition or more recently 
giant magnetoresistance effects in magnetic multilayers.

\vspace{10pt}

\hspace{\parindent}The RKKY 
interaction is generically given by $
{\cal I}_{RKKY}(r_{i},r_{j},E)=J^{2}\chi(r_{i},r_{j},E) 
{\bf S}_{r_{i}}.{\bf S}_{r_{j}}$ where J is the interaction between
the localized moment ${\bf S}_{r_{i}}$  and the spin of the itinerant
electrons, and $\chi(r_{i},r_{j},E)$
contains the sum of all the electron-hole propagation paths from
 $|r_{i}\>$ to $|r_{j}\>$. The susceptibility can
be written down as

\vspace{5pt}

$$
\chi(r_{i},r_{j})= 2 \Re e {\displaystyle \int_{E>E_{F}\atop
E'<E_{F}}}dEdE' {{\<r_{i}\mid \delta (E-{\cal H})\mid
r_{j}\>\<r_{j}\mid \delta(E'-{\cal H})\mid r_{i}\>}\over{E-E'}}
$$

\noindent
and by development of spectral measure on Chebyshev polynomials 
one gets~\cite{RO-RKKY,ROB-RKKY}

\begin{eqnarray}
\chi_{ij}&=&2\  \Re e {\displaystyle\sum_{m,n}} \ I_{mn} 
\<r_{i}\mid P_{n}({\cal H})\mid r_{j}\> \<r_{j}\mid P_{m}({\cal
H})\mid r_{i}\>\nonumber \\
I_{mn}&=&{\displaystyle\int_{E>E_{F}\atop E'<E_{F}}}N(E)N(E')
{{P_{m}(E)P_{n}(E')}\over{E-E'}}\ dEdE' \nonumber
\end{eqnarray}

\noindent
where the coefficients $I_{mn}$ have been previously
 defined for the Hall conductance.

\vspace{10pt}

\hspace{\parindent}In metallic systems,
 the interaction is calculated exactly
${\cal I}_{RKKY}(r,E)\sim A({\bf r})\cos(2k_{F}r+\delta({\bf
r}))/r^{3}$, contrary to quasiperiodic or disorder systems where
there is no simple analytical form. However, 
in weakly disordered systems, one can evaluate the quantum fluctuations that arise in
 the higher moments of the interaction.~\cite{RKKY-Meso} It is found
that only even moments lead to significant contributions : 

\begin{eqnarray}
\<\chi^{2p}(\mid r_{i}-r_{j}\mid)\>\ &\simeq& \ \Omega_{p}\biggl 
( {{\rho^{2}(E_{F})}\over{\mid r_{i}-r_{j}\mid^{2d}}}\biggr)^{p}\sim 
\bigl(\<\chi^{2}(\mid r_{i}-r_{j}\mid)\>\bigr)^{p}\nonumber  \\
\<\chi^{2p+1}(\mid r_{i}-r_{j}\mid)\>\ &\simeq&  \ 
\hbox{exp}(-\frac{\mid r_{i}-r_{j}\mid}{\strut l_{m}}) \nonumber
\end{eqnarray}

\noindent
with $\rho(E_{F})$ the DoS at Fermi level, $l_{m}$ the mean free
path, whereas $\Omega_{p}$ is constant independent of the parameter of 
the hamiltonian.

%%%%%%%%  FIGURE V AND VI  %%%%%%%%%%%%%%%%%%%%%%%%%%%%%%%%%%%%%%%

\vspace{10pt}

\hspace{\parindent}Quasiperiodic systems
 cannot be described by such averaging process, so
that the use of recursion method gives here some interesting 
quantitative informations.~\cite{RO-RKKY,ROB-RKKY} On Fig. 7
and 8, the TDoS for a 2D quasiperiodic Fibonacci quasilattice, as
well as typical signature of aperiodic 
long range order are depicted. The strength of the quasiperiodic
potential is $V_{qp}=0.4t$ (with t the constant hopping integral between first neighbors)
and the susceptibility is given in a-units, with a the lattice
spacing. In the Fig. 8, typical features of
the interaction are shown, but due to the 
complicated nature of quasiperiodic potential, no Fermi wavelength can 
be properly defined and oscillations exhibit resurgences that are
absent from the periodic potential.~\cite{ROB-RKKY}

\vspace{10pt}

\section{Possible applications for large scale computational methods}

\vspace{10pt}

\hspace{\parindent}The expansion of any operator on orthogonal
 polynomials can also be applied in the context of large scale
 computational methods, which aim at reducing large memory and CPU time
costs, for investigating more realistic models. For instance, spectral
properties and optical spectra for realistic model of Silicon quantum dots
 have been evaluated and quantum confinement investigated.~\cite{Wang-I} One can also
consider thermodynamical properties of quantum systems thanks to the
development of the partition functions ${\cal
Z}(\beta)=\hbox{Tr}[e^{-\beta{\cal H}}]=\sum_{n}(\int dEe^{-\beta
E}{\cal P}_{n}(E))\times {\cal P}_{n}({\cal H})$, with $\{ {\cal P}_{n}\}$ a
suitable basis of orthogonal polynomials. Indeed, one needs to consider average 
quantities as $\langle \hat{\cal A}\rangle =
 Tr [\hat{\rho} \hat{\cal A}]/Tr[\hat{\rho}]$ ($\hat{\rho}$ the
density operator), but if the expansion of the partition function can
be easily done in the case of a scalar argument,
 when applied to an operator (for large scale systems), 
one may not be able to switch into the eigen-representation of the
Hamiltonian since full diagonalization is practically 
impossible.~\cite{FD-I,FD-II,DOS-I} Consequently, polynomial expansions of 
operator are the only available alternative technique.

\vspace{10pt}

\hspace{\parindent}Finally, in order to perform realistic atomistic
 calculations of binding energies and interatomic
forces, real-space methods turn out to be efficient and accurate
O(N)-methods.~\cite{Petitfor}

\vspace{10pt}

\section{Conclusion}

\vspace{10pt}

\hspace{\parindent}Results on quantum transport by means of a
real-space methods have been presented. The generality of the use of orthogonal
polynomials for high-dimensional non periodic systems has been
discussed. In two-dimensional disordered systems with magnetic field,
global phase diagram and new universalities of the quantum Hall effect 
may be studied by this method. One notes that, nowadays, lots of
efforts are devoted to improve the efficiency of 
available numerical algorithms. The well-known
Car-Parinello method has opened new perspectives for ab-initio
electronic structure calculations, but the development of the so-called ``order-N
scheme''~\cite{Fujiwara} is of a major importance to investigate complex
molecular systems. Quantum transport has also opened new challenges
since the development of nanostructures (quantum dots, nanotubes,..)
has become crucial in the field of electronic devices. 
In this context, further ``order-N'' schemes to
investigate electronic propagation also need to be steadily improved.

\vspace{10pt}

\section{Appendix on orthogonal polynomials}

\vspace{10pt}

\hspace{\parindent}
The principle of the real-space algorithms, used in this article, 
is based on expansion of spectral measures on orthogonal 
polynomials. Let us mention the requirements as well as the key points of this approach. 
For any $\rho(E)$ defined by
$\int_{a}^{b} \rho(E)dE=1$, there exists a family of
 orthogonal polynomials ${\cal P}_{n}(E)$ of degree n, such that

\begin{eqnarray}
&\int_{-\infty}^{+\infty}& N(E){\cal P}_{n}(E){\cal P}_{m}(E)dE=\delta_{nm}\nonumber\\
&\rho(E)& \sum_{n}{\cal P}_{n}(E){\cal P}_{n}(E')=\delta(E-E')\nonumber
\end{eqnarray}

\noindent
for $E,E'$ belonging to the spectral subset of $\rho(E)$ and
$\rho(E)\ne \rho(E')$. These polynomials generally satisfy a
three-term recursive relation 
$EP_{n}(E)=a_{n}{\cal P}_{n}(E)+b_{n}{\cal P}_{n+1}(E)+b_{n-1}{\cal P}_{n-1}(E)$
with $b_{-1}=0, n\geq 0$, $a_{n}, b_{n}$ related to the moments of 
spectral function.~\cite{Gaspard,Szego} However, the expansion of
spectral measure can be also done in a
arbitrary basis of polynomials, for instance on Chebyshev polynomials ${\cal Q}_{n}$ :

\begin{eqnarray}
\delta(E-{\cal H})&=&\tilde\rho(E)\sum_{n} 
{\cal Q}_{n}(E){\cal Q}_{n}({\cal H})\nonumber\\
&=&\rho(E)\sum_{n} {\cal P}_{n}(E){\cal P}_{n}({\cal H})\nonumber
\end{eqnarray}

\noindent
where  $\tilde\rho(E)=1/ \pi\sqrt{
4b^{2}-(E-a)^{2}}$ and $\rho(E)$ the total density of states. 
Here, $a$ and $b$ define the band edges and bandwidth of the
spectral function associated to Chebyshev polynomials.~\cite{Magnus}
Green's functions can be obtained by 
Hilbert transformation of $\rho(E)$ (for absolute continuous spectrum) :

\begin{eqnarray}
 G(z)&=&\int dE\frac{\rho (E)}{(z-E)} \nonumber\\ 
\rho(E)&=&\lim_{\eta\to 0^{+}}\ -\frac{1}{\pi} \Im m  G(E+i\eta) \nonumber
\end{eqnarray}

\noindent
and general analytic properties of the Green's function can be investigated 
through the properties of the considered polynomials.

\section{\sc Acknowledgments}
 The author would like to thanks to  T. Ando and D. Mayou for useful comments. He 
is also indebted to the European Commission and the Japanese
 Society for Promotion of Science (JSPS) for joint financial support 
(Contract ERIC17CT960010), and to T. Fujiwara from Department of
Applied Physics of Tokyo University for his kind hospitality.

\vfill\eject

\section{Figures captions}

\vspace{20pt}

\noindent
Fig. 1. Recursion coefficients $b_{n}$ and corresponding total density
of states for three different strengths of the magnetic field.

\vspace{20pt}

\noindent
Fig. 2.  Diagonal conductivity for quantum Hall systems with different 
disorder strengths ($\alpha=1/3, \eta=0.15$) in $e^{2}/h$ unit.

\vspace{20pt}

\noindent
Fig. 3. Diagonal conductivity versus Fermi energy and finite imaginary part
of the Green's function $\eta$ (inset).

\vspace{20pt}

\noindent
Fig. 4. Diagonal conductivity versus Fermi energy and
disorder strength for $\alpha=1/16$.

\vspace{20pt}

\noindent
Fig. 5. Dissipative conductivity as a function of disorder strength. Stars 
are associated with a Fermi level in the gap of the pure system (no
disorder) whereas Fermi level lies within the middle of a Landau band
for diamonds.

\vspace{20pt}

\noindent
Fig. 6. Components of the dissipative conductivity as a function of disorder strength
and Fermi energy (black curves), and averaged result (red curve).

\vspace{20pt}

\noindent
Fig. 7. Total density of states for a two-dimensional quasiperiodic
lattice ($V_{qp}$ is the strength of the Fibonacci quasiperiodic
potential). In the onset is given the energy interval for which
corresponding susceptibility are calculated.

\vspace{20pt}

\noindent
Fig. 8. Electronic susceptibility (in a-units, a lattice parameter)
 as a function of Fermi energy with
1.a,2.a,2.b,2.c,3.a,3.b,3.c corresponding respectively to Fermi energy
$-1.9+0.1\times \lambda,\ \lambda=0,1,2,3,4,5,6$.

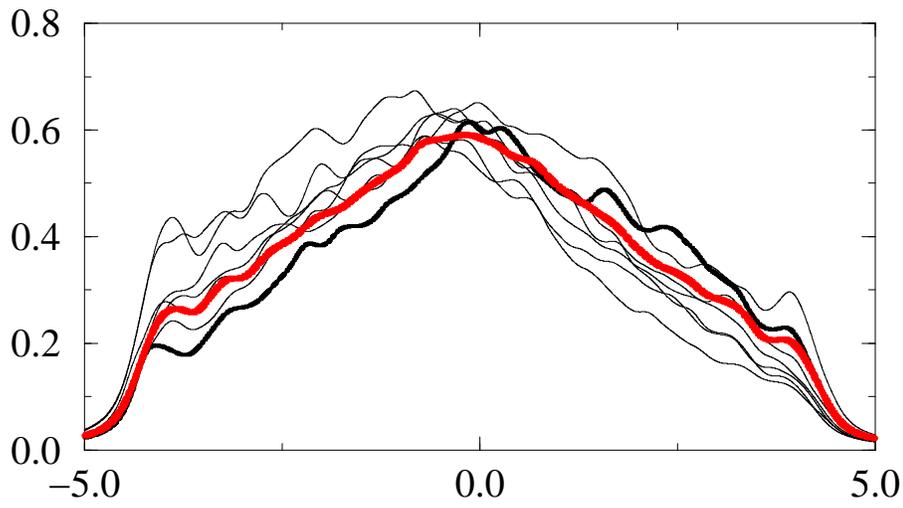
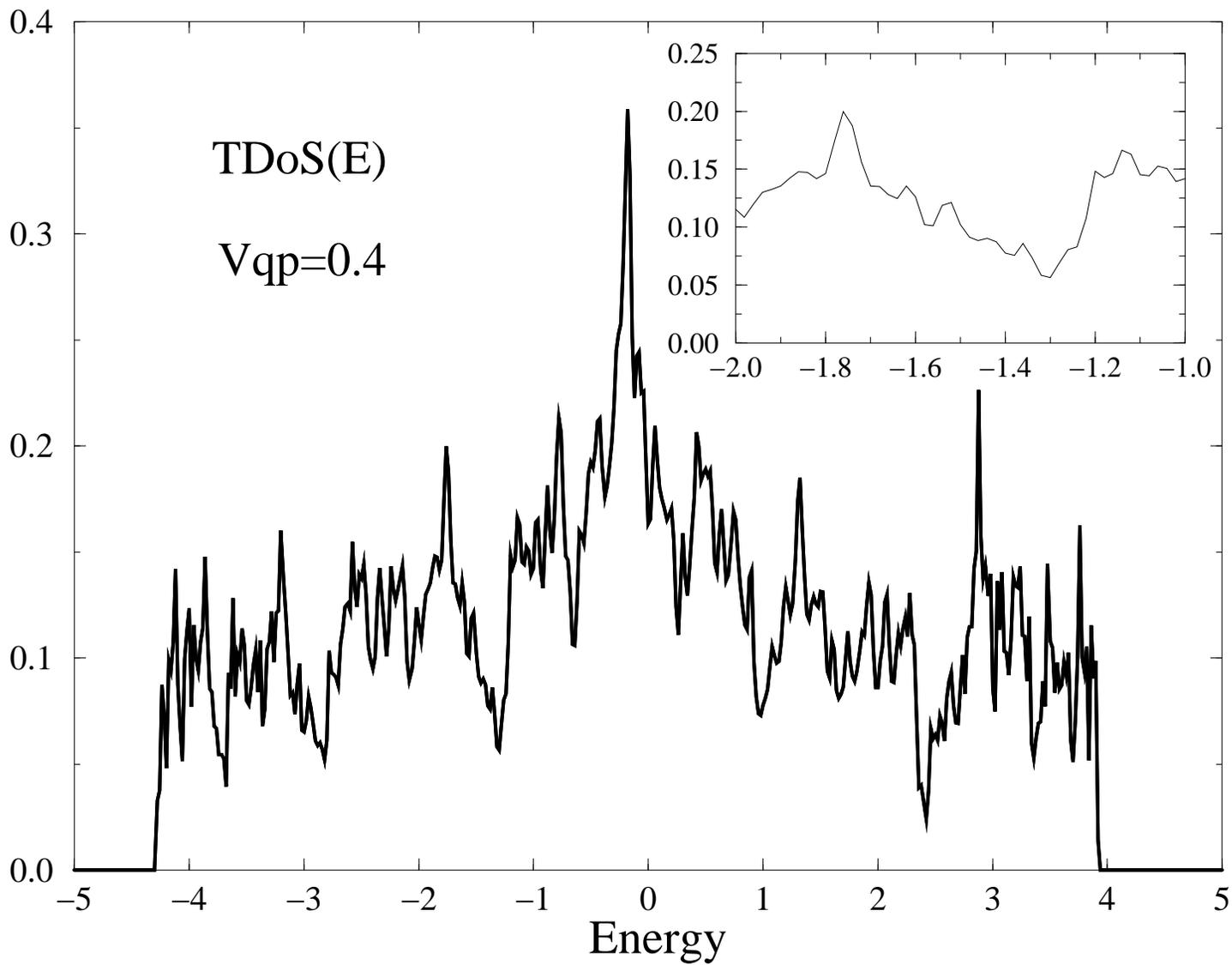
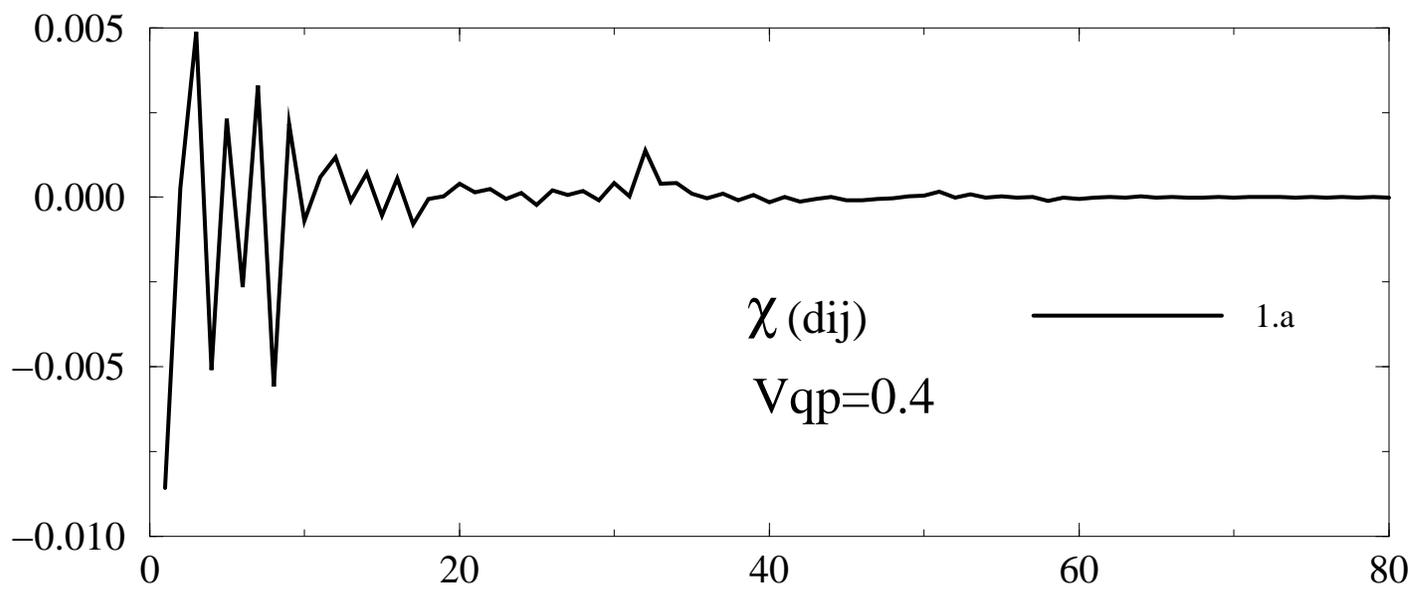
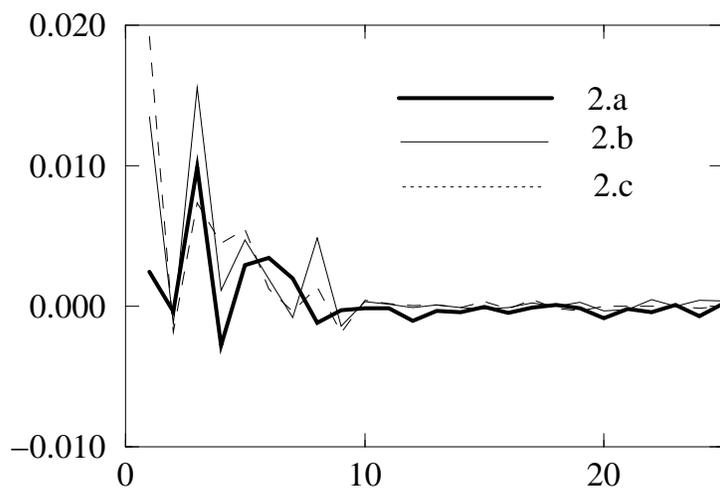
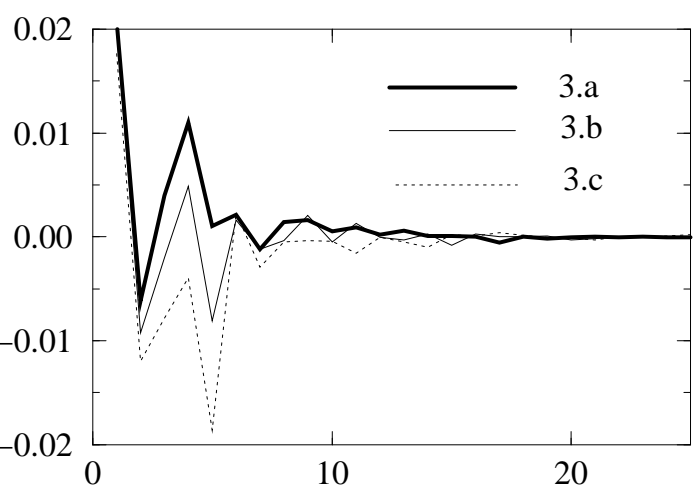
\end{document}